\documentstyle[prd,aps]{revtex}
\begin{document}
\draft
\twocolumn[\hsize\textwidth\columnwidth\hsize\csname
@twocolumnfalse\endcsname
\title{ Reply to: Comments on ``Analytic and Numerical Study of Preheating Dynamics''}

\vskip 1cm
\author{{\bf D. Boyanovsky$^{(a)}$, 
H.J. de Vega$^{(b)}$ and  R. Holman$^{(c)}$ }}   
\address
{ (a)  Department of Physics and Astronomy, University of
Pittsburgh, Pittsburgh, PA. 15260, U.S.A. \\
 (b)  LPTHE, \footnote{Laboratoire Associ\'{e} au CNRS UA280.}
Universit\'e Pierre et Marie Curie (Paris VI) 
et Denis Diderot  (Paris VII), Tour 16, 1er. \'etage, 4, Place Jussieu
75252 Paris, Cedex 05, France \\
 (c) Department of Physics, Carnegie Mellon University, Pittsburgh,
PA. 15213, U. S. A. }
\address{}
\date{August 1996}
\maketitle
\begin{abstract}
Reply to the comment by L. Kofman, A. Linde and A.A. Starobinsky
 (hep-ph/9608341) to 
 our article ``Analytic and Numerical Study of Preheating Dynamics''
(hep-ph/9608205).

\end{abstract}
 \vskip2pc]

Preheating is the stage of profuse production of out of equilibrium particles
either by parametric amplification of quantum fluctuations or by the growth of
spinodal instabilities prior to the reheating stage in inflationary
cosmologies. Such a mechanism was recognized to be more efficient than single
particle decay\cite{bran,kls} in terms of reheating the post inflationary
universe.

The origin of such a mechanism was traced to parametric amplification of
quantum fluctuations by the evolution of the inflaton zero mode. In a previous
article\cite{disi} we studied such processes, offering a detailed numerical
analysis in several approximation schemes. We found some very interesting new
phenomena, especially in the case of broken symmetry, where, for ``slow-roll''
initial conditions, the zero mode is driven back to the origin and most of the
energy has been transferred to the non-equilibrium fluctuations. The
quantitative features, such as time scales for the preheating stage, the rate
of particle production and the total number of particles produced are strongly
model dependent.

The possibility of symmetry restoration via this non-equilibrium fluctuations
was put forth in reference\cite{restor}, wherein such a mechanism was offered
as an explanation for the unusual behavior found in\cite{disi}. In a recent
article\cite{big} we provided a more thorough analysis of these new phenomena,
focusing on the description in Minkowski space-time, within the large
$ N $
approximation. This approximation provides a consistent, renormalizable, energy
conserving scheme that permits an accurate numerical analysis of the evolution
equations. 

In this article we recognized that approximate schemes using the Mathieu
equation to describe the stage of parametric amplification, though
qualitatively correct in the sense that a solid intuition on the processes can
be obtained from such an approximation, are {\em quantitatively} in error.  We
also recognized that the possibility of symmetry restoration is not borne out
in the broken symmetry phase, ruling out the potential explanations
proposed in ref.\cite{restor}
for our previous results. In our recent article we have provided an analytic
and numerical study for this phenomenon and explained its physical reasons.

In a recent comment to our article \cite{big}, L. Kofman, A. Linde and
A. A. Starobinsky (KLS)
present a criticism of our results. Which we analyze within the context of our
article below.

\section{The Setting:}

As mentioned above, the phenomenon of preheating and reheating is strongly
model dependent, as well as highly non-perturbative in nature. Our study
focuses on the description of this phenomenon in {\em Minkowski space} to
provide a detailed and deeper understanding within a simpler setting.  The
non-perturbative nature of this phenomenon requires that a consistent
approximation scheme be invoked, and we used the large $ N $ expansion
which is a 
non-perturbative scheme that can be consistently improved, unlike the Hartree
approximation.  

In this approximation, the dynamics, including backreaction effects, can be
summarized in the evolution equations for the expectation value of the zero
mode and the fluctuations. We analyzed two different cases: unbroken and broken
symmetry.  The comment by KLS introduces expansion the of the universe, which
certainly does not apply to our study.

\subsection{\bf Unbroken symmetry: Lam\'e vs. Mathieu}
In the unbroken symmetry case we examined the case of very large
energy density initially stored in the zero mode. In the large N limit the
field expectation value $\phi(t)= <\Phi(\vec x,t)>$ (with $<\cdots >$ being the
expectation value in the translational invariant but non-equilibrium quantum
state) obeys the evolution equation

\begin{equation}
\ddot{\phi}(t)+M^2[\Phi]\; \phi(t) = 0  \label{zeromode},
\end{equation}
while the mode functions driving the quantum fluctuations
obey:
\begin{equation}\label{modos}
{\ddot \chi}_k   + \left( {k^2} +
M^2[\Phi] \right)  \chi_k(t) = 0
\end{equation}

The effective ``mass'' term $M^2[\Phi]$ includes the evolution of the zero
mode and the back-reaction effects of the mode themselves through the quantum
fluctuations. Therefore this is a highly non-linear integro-differential
problem. We emphasized that the full backreaction problem is energy conserving
and provided a detailed account of the renormalization aspects. Note that none
of these issues were taken up by KLS in their analysis.

In the initial stage, all the energy is assumed to be in the zero mode of the
field. This corresponds to choosing as initial state the ``vacuum'' for the $k
\neq 0$ modes. For very weak coupling the back-reaction effect will be
negligible up to a time we call the preheating time; we will estimate this time
below. Thus for the early stages of the dynamics and for very weak coupling one
can approximate

\begin{equation} \label{prehM}
M^2[\Phi] \simeq m^2 + { {\lambda} \over 2} \phi(t)^2
\end{equation}

Now, using the classical oscillating behaviour of $ \phi(t) $ one is led by
eq.(\ref{modos}) to an effective mass that oscillates in time. In this
approximation (which is indeed very good for weak coupling and during the
preheating stage, see \cite{big}), eq.(\ref{modos}) exhibits {\em parametric
resonance} as noticed first in ref\cite{bran}. Namely, there are allowed and
forbidden bands in $ k^2 $. The modes within the forbidden bands grow
exponentially, whereas those in the allowed bands remain bounded in absolute
value. The growth of the modes in the forbidden bands is interpreted as the
production of particles are with these momenta and the production rate is
determined by the imaginary part of the Floquet index.  This exponential,
non-perturbative growth of quantum fluctuations will lead to strong
backreaction effects once the contribution from the quantum fluctuations to
$M^2[\Phi]$ becomes of the same order as the tree-level terms. When this
happens the approximation (\ref{prehM}) breaks down, and the energy transferred
from the zero mode to the produced particles is a large fraction of the initial
zero mode energy.

In approximations that do not include the backreaction effects there is
infinite particle production, since the effect of draining energy from the zero
mode is not taken into account. The full backreaction problem and the
approximation used in our work maintains energy conservation, and thus displays
the shutting down of the particle production when the back-reaction becomes
important; this is the end of the preheating stage.

Now, in order to compute quantitatively the number of particles produced, the
behavior of the zero mode is required. In ref.\cite{bran,kls} $ \phi(t) $ is
approximated by a cosine in the calculations. The mode equations then become
the Mathieu equation. As shown in ref.\cite{big}, the exact classical
solution is actually a cn Jacobi function. The difference between this and a
trigonometric function is profound.

Let us now compare the results from the exact mode solutions obtained in
ref.\cite{big} with the Mathieu equation approximation to it.  In units where $
m^2 = 1 $ and setting $ \eta(t) \equiv \sqrt{ {\lambda} \over 2} \; \phi(t) $,
one finds
\begin{eqnarray}\label{etac}
\eta(t) &=& \eta_0\; \mbox{cn}\left(t\sqrt{1+\eta_0^2}, {\bar k}\right)
\cr \cr
 {\bar k} &=& {{\eta_0}\over{\sqrt{2( 1 +  \eta_0^2)}}}\; , 
\end{eqnarray}
where cn stands for the Jacobi cosine and we choose for initial
conditions $ \eta(0) =  \eta_0\; , \; {\dot \eta}(0) = 0 $.

Inserting this form for $\eta(\tau)$ in eqs.(\ref{prehM}) and (\ref{modos})
yields
\begin{equation}\label{modsn}
 \left[\;\frac{d^2}{dt^2}+k^2+1+   \eta_0^2\;
\mbox{cn}^2\left(t\sqrt{1+\eta_0^2}, {\bar k}\right) \;\right]
 \chi_k(t) =0 \; . \label{nobackreaction}
\end{equation}
This is the Lam\'e equation for a particular value of the coefficients that
make it solvable in terms of Jacobi functions (see\cite{big} and references
therein).  As shown in ref.\cite{big}, this equation has only {\em one}
forbidden band for positive $ k^2 $, which runs from $ k^2 = 0 $ to $ k^2 =
{{\eta_0^2}\over 2} $. One can choose Floquet solutions of
eq.(\ref{nobackreaction}) fulfiling the relation
\begin{equation}\label{floq}
 U_k(t + 2  \omega) =   e^{i F(k)} \; U_k(t),
\end{equation}
where the Floquet indices $ F(k) $ are independent of $t$. In the
forbidden band the  $ F(k) $ posses an imaginary part. Their exact
form results \cite{big}
$$
 F(k) = -2 i K( {\bar k}) \;  Z(2  K( {\bar k}) \,v)  +  \pi
$$
where $ Z(u) $ is the Jacobi zeta function (see\cite{big} and references
therein) and $ v $ is a function of $ k $ in the forbidden band defined by
\begin{equation}\label{qprohi}
k =  {{\eta_0}\over {\sqrt2}}\, \mbox{cn}(2  K( {\bar k})\,v,k) \; ,
\; 0 \leq v \leq \frac{1}{2}. 
\end{equation}
All these elliptic functions posses fastly convergent expansions in
powers of the elliptic nome 
$$
q \equiv  e^{-\pi K'( {\bar k})/ K( {\bar k})} \; .
$$ 
Since $ 0 \leq {\bar k}  \leq 1/\sqrt2 $ [see eq.(\ref{etac})], we have
\begin{equation}\label{qsomb}
 0 \leq  q  \leq e^{-\pi} = 0.0432139\ldots \; . 
\end{equation}
Then, 
\begin{equation}
F(k) = 4i\, \pi  \;  q \; \sin(2\pi v)\;\left[ 1 + 2 \, q
\; \cos2\pi v   + O( q^2)\right]  + \pi \; .\label{easyfloquet}
\end{equation}
The imaginary part of this function has a maximum at $ k = k_1 =
 \frac12 \; \eta_0 \; (1 -  q )  + O( q^2)  $ where \cite{big}
\begin{equation}\label{Flame}
{\cal F} \equiv  Im F(k_1) = 4\, \pi \; q + O(q^3) \; .
\end{equation}
This simple formula gives the maximun of the imaginary part of the Floquet
index in the forbidden band with a precision better than $ 8.\, 10^{-5} $. $ q
$ can be expressed in terms of $ \eta_0 $ as follows \cite{big}
$$
q  =  \frac12 \;  {{ (1+\eta_0^2)^{1/4} -  (1+\eta_0^2/2)^{1/4}}
\over { (1+\eta_0^2)^{1/4} +  (1+\eta_0^2/2)^{1/4}}}  \; .
$$
with an error smaller than $\sim 10^{-7} $.

Let us now proceed to the Mathieu equation analysis of this problem. The cn
Jacobi function can be expanded as (see \cite{big} and references therein)
$$
{\rm cn} (z, {\bar k}) = (1-q) \cos(1-4q)z + q \cos3z + O(q^2) \; .
$$
To zeroth order in $ q $ we have
$$
\eta(t)^2 = {{\eta_0^2}\over 2} \left[ 1 + \cos(2t\sqrt{1 + \eta_0^2})
\right] + O(q) \; .
$$
and $ 2 \omega = \pi/\sqrt{1 + \eta_0^2} + O(q) $.  Under such approximations
eq.(\ref{modsn}) becomes the Mathieu equation \cite{abr}
\begin{equation}\label{mathieu}
{{d^2 y}\over {dz^2}} + \left( a - 2  {\bar q} \cos2z \right)y(z) = 0
\; ,
\end{equation}
where
$$
a = 1 + {{k^2 -  {{\eta_0^2}\over 2} }\over {\eta_0^2 + 1 }} \; , \; 
 {\bar q} = {{\eta_0^2}\over{ 4(\eta_0^2  + 1)}}  
$$
and $ z = \sqrt{ \eta_0^2 + 1 } \; t $. Notice that $ 0 \leq   {\bar q}
\leq 1/4 $ in the present case. Eq.(\ref{mathieu}) posses an infinite
number of forbidden bands for $ k^2 > 0 $. The lower and upper band edges for the first
band are given by \cite{abr}
$$
k^2_{inf} =   {{\eta_0^2}\over 4}\left[ 1 -  {{\eta_0^2}\over{
2^5(\eta_0^2 + 1)}}  +  {{\eta_0^4}\over{ 2^{10}(\eta_0^2 + 1)}} + \ldots
\right] \; ,
$$

$$
k^2_{sup} =   {{\eta_0^2}\over 4}\left[ 3 -  {{\eta_0^2}\over{
2^5(\eta_0^2 + 1)}}  -  {{\eta_0^4}\over{ 2^{10}(\eta_0^2 + 1)}} + \ldots
\right] \; .
$$
These values must be compared with the exact result for the Lam\'e equation
given by (\ref{nobackreaction}) : $ k^2_{inf} = 0 \; , \; k^2_{sup} =
{{\eta_0^2}\over 2} $.  Although the width of the band is well approximated by
the Mathieu equation, its absolute position is not. The numerical values of the
maximum of the imaginary part of the Floquet index are given in Table I and
compared with the exact values from eq.(\ref{Flame}).

\begin{table} \centering
\begin{tabular}{|l|l|l|l|}\hline
$ \eta_0 $ &  $ {\cal F}_{Lame} $  &  $ {\cal F}_{ Mathieu} $ &  
 $ \% error $ \\ \hline 
$ $&  $ $ & $ $ & $ $ \\
1 & $ 0.2258 \ldots $ & $ 0.20 \ldots $ & $ 13$\% \\
$ $&  $ $ & $ $ & $ $ \\ \hline
$ $&  $ $ & $ $  & $ $ \\
$ 4 $ & $ 0.4985\ldots $ & $ 0.37\ldots $ & $ 35$\% \\
$ $&  $ $ & $ $ & $ $ \\ \hline
$ $&  $ $ & $ $  & $ $ \\
 $ \eta_0 \to \infty $ &  $ 4\pi e^{-\pi} = 0.5430\ldots $ & $ 0.39\ldots
$ & $ 39$\% \\ $ $&  $ $ & $ $ & $ $ \\ \hline
\end{tabular}

\bigskip

\label{table1}
\caption{ The maximum of the imaginary part of the Floquet index ${\cal
F}$ for the Lam\'e  equation and for its Mathieu approximation.}
\end{table}

We see that the Mathieu approximation {\em underestimates} the exact result by
a fraction ranging from $13 $ \% to $ 39$\%. The second forbidden band in the
Mathieu equation yields $ {\cal F}_{ Mathieu} = 0.086\ldots $ for $ \eta_0 \to
\infty $. This must be compared with $ {\cal F}_{Lame} = 0 $ corresponding to
the fact that there is only one forbidden band in the Lam\'e equation.

In ref.\cite{kofman}, a large
discrepancy between Lame and Mathieu
Floquet indices has been reported within a different approximation scheme.

It is worth mentioning that
 differences in the Floquet indices such as those
displayed , enter in the exponent. In the large N approximation we see that the
discrepancy between Mathieu and Lame is very large and therefore cause very large errors in
{\bf quantitative} estimates of particle production and preheating time. 

Although in some cases as found by KLS the Mathieu equation gives a reasonable
estimate, in some other cases it clearly does not. This was the point raised in our
article, each case must be treated in its own right. 

For example, the number of particles produced during reheating is of the order
of the exponential of $ 2 {\cal F} $ times the reheating time in units of $
\pi/\sqrt{ 1 + {\eta_0^2}} $. An error of $25$\% in $ {\cal F} $ means an error
of $25$\% in the exponent, so that one would find $10^9$ instead of
$10^{12}$.
 
The Mathieu equation approximation would be  exact in absence of the $
\lambda \Phi^4 $ inflaton self-coupling. That is, for the classical
potential\cite{kls}
\begin{equation}\label{Vines}
V = \frac12 m^2 \Phi^2 +  g  \sigma^2 \Phi^2 \; ,
\end{equation}
one can consider as classical solution $  \Phi(t) =  \Phi_0 \;
\cos(mt) , \; \sigma = 0 $. However, the potential (\ref{Vines}) is
unstable under renormalization (a $ \Phi^4 $ counterterm is needed from
the one-loop level). Hence, the $ \lambda = 0 $ choice is a fine-tuning not
protected by any symmetry.

The mode equations (\ref{modos}) apply to the selfcoupled $ \lambda \; \Phi^4 $
scalar field. Models for reheating usually contain at least two fields: the
inflaton and a lighter field $ \sigma(x) $ in which the inflaton decays. For a
$ g \, \sigma^2 \, \Phi^2 $ coupling, the mode equations for the $
\sigma $ field take 
the form\cite{kls,disi,big} (adapted to Minkowski space which is
the focus of our article) 
\begin{equation}\label{modsi}
{\ddot V}_k  +  \left( {k^2} +
m_{\sigma}^2 + {g\over {\lambda}} F[\eta(.)] \right)  V_k(t) = 0
\end{equation}
A new dimensionless parameter $ { g \over {\lambda}} $ appears here. Neglecting
the $ \sigma $ and $ \Phi $ backreaction, we have
\begin{equation}\label{preH}
 F[\eta(.)] \simeq   \eta^2(t) \; .
\end{equation}
In
ref.\cite{disi,big}, we show 
that abundant particle production (appropriate for reheating) occurs even for
$ g = \lambda$.

In hep-ph/9608341 it is stated that
``the main subject of investigation of   \cite{big} was
the theory $m^2\phi^2/2 +\lambda\phi^4/4$.  There are two main regimes there:
$\phi \ll m/\sqrt \lambda$ and  $\phi \gg m/\sqrt \lambda$. The authors of
\cite{big} do not make any distinction between these two regimes because
they neglect expansion of the universe. However, from  \cite{kls} it
follows that  in expanding universe there is no parametric resonance at all for
$\phi \ll 
m/\sqrt \lambda$. Therefore all results of   \cite{big} related to
parametric resonance in this 
regime do not give a correct description of reheating in the theory
$m^2\phi^2/2 +\lambda\phi^4/4$.''

Firstly, parametric resonance and universe expansion are independent
processes that desserve separate investigation. Secondly, in our
investigations in Minkowski spacetime \cite{big} we provide analytic
expressions which hold both for small and large amplitudes. One
explicitly sees there that the parametric resonance dies in the limit
of  small amplitude.

Moreover, in ref.\cite{marco} we studied the small amplitud case
$\eta <<1$, 
in Minkowski space and compared explicitly with the one-loop and linear
relaxation (single particle decay) results showing in detail how
parametric amplification 
for large amplitudes merges with linear relaxation at small amplitudes. 

Therefore, the above statement in  hep-ph/9608341 cannot be applied
to  ref.\cite{big} neither to our previous works.

Eqs.(\ref{modsi})-(\ref{preH}) become Lam\'e equations when $ \eta(t) $ is
approximated by the classical solution in Minkowski spacetime given by
(\ref{etac}). This Lam\'e equation is solvable in closed form when the
couplings $ g $ and $ \lambda $ are related as follows
$$
 {{2 g}\over {\lambda}} = n(n+1) \; , \; n=1,2,3,\ldots
$$
In those cases there are $ n $ forbidden bands for $ k^2 \geq 0 $. The Lam\'e
equation exhibits an infinite number of forbidden bands for generic values of $
{g\over {\lambda}} $. The Mathieu and WKB approximations have also been applied in
the non-exactly solvable cases \cite{kls}. However, as the above analysis shows,
(see Table I) these results cannot be trusted quantitatively. The only
available precise method consists on accurate numerical calculations as those
of ref.\cite{disi,big} (where the precision was at least $ 10^{-6} $).
  
As soon as the quantum fluctuations grow and cease to be negligible compared
with the the classical contribution (\ref{preH}), all the approximations
discussed so far (Lam\'e, Mathieu, etc.) break down. This time is the so-called
preheating time $ t_{reh} $ \cite{big}.  One can estimate $ t_{reh} $ by
equating the zero mode energy (\ref{preH}) with the estimation of the quantum
fluctuations derived from the unstable Floquet modes \cite{kls,big}. Such
estimation yields accurate estimates when the Lam\'e Floquet indices are used
\cite{big}. However, as emphasized before, because of the non-perturbative
nature of this time scale $ \approx \ln[1/\lambda]$\cite{big}, differences in
the Floquet indices stemming from a Mathieu equation approximation to a Lam\'e
equation lead to severe errors in the quantitative estimate of such a time
scale.

Although a fairly accurate estimate of $t_{reh}$ can be obtained via the Lam\'e
equation, with the exact classical solution for the zero mode, in order to
compute physical magnitudes beyond $ t_{reh} $, one {\bf must} solve
self-consistently the field equations including the back reaction. Clearly this
requires a numerical treatment.  In ref.\cite{disi,big} this is done for the $
N \to \infty $ limit and in ref.\cite{disi} to one-loop order.  Such a study
led to a very clear physical description of the non-equilibrium gas of created
particles, and its equation of state\cite{big}.

Thus, at the expense of re-iterating our conclusions and comments in our
article, a Mathieu equation approximation provides a {\bf qualitative,
intuitive} description of parametric amplification, and the non-perturbative
phenomena associated with particle production out of equilibrium.  However, for
a {\bf quantitative and trustworthy} estimate of the physical time scales and
production rates one {\bf must} study the proper problem in its full
complexity. Differences in Floquet indices, propagate exponentially making the
qualitative estimates based on Mathieu unreliable whenever the Mathieu equation
is not the proper description. Ultimately one must resort to a full numerical
scheme to study the dynamics at long times, making sure that the approximations
involved maintain energy conservation (or covariant conservation).  Given that
this phenomenon is not universal and strongly model dependent, one equation
{\em does not fit all} the different situations and a careful and consistent
analysis of each particular scenario is needed.

\subsection{Broken Symmetry:} 

Our study of the dynamics in the broken symmetry case in connection with the
possibility of symmetry restoration as advocated by KLS was sparked by the
following statement in their article ``Nonthermal Phase
Transitions...'' by KLS, 
Phys. Rev. Lett. 76, 1011, (1996)\cite{restor} and again by Kofman\cite{kofman}
in his review article.  In particular, on page 1012,
second column, third paragraph of this article: ``The mechanism of symmetry
restoration described above is very general; in particular, it explains a
surprising behavior of oscillations of the scalar field found numerically in
the $O(N)$-symmetric model of Ref.[6]''.  This reference is to our previous
article\cite{disi}, and there we studied the same situation that we re-analyzed
in our latest article: Minkowski space-time, large $ N $ and broken
symmetry phase, 
with the zero mode of the field beginning very close to the origin, i.e. with
``slow-roll'' initial conditions. It is precisely this situation that we
studied in deeper detail in our latest article and concluded after a thorough
analytic and numerical study that there is no symmetry restoration by the
quantum fluctuations. Energy is conserved and the sum
rule
\begin{equation}
-1+\eta(\infty)^2+g\Sigma(\infty)=0
\end{equation}
which is nothing but a statement of massless pions and a consequence of
Goldstone's theorem, is satisfied. The final value $\eta(\infty)$ is obtained
dynamically, and depends on the initial condition that determines the total
energy of the system which is conserved by the numerics to one part in
$10^{10}$ which is our numerical accuracy.

We have provided exhaustive evidence for this behavior both analytically and
numerically and showed unequivocally that there is {\bf no
possibility} of symmetry 
restoration as envisaged by KLS and as quoted by\cite{restor}
  in the situation that we studied.

We have learned privately that Cooper, Habib, Mottola and Kluger\cite{cooper}
have obtained similar conclusions in their thorough and independent study of
the broken symmetry phase, within the same approximation scheme. Their results
are complementary to ours, Cooper et. al. study the strong coupling regime, but
the behavior is consistent with the results obtained for the weak coupling case
in our article. 

In their comment, KLS seem to agree with our conclusion that in this particular
situation studied in our articles, when the expectation value is
 released near the potential hill, symmetry restoration does not occur. We have
pointed out in\cite{disi} and more recently in\cite{big} that the notion of the
effective potential is irrelevant for the dynamics and these effects should be
understood as a dynamical change of the effective action. The minima of the
effective action attained dynamically are very different from those of the effective potential.

In the situation of  `chaotic initial conditions' but with a broken
symmetry tree level potential, the 
issue of symmetry breaking is more subtle. In this case the zero mode
is initially displaced with a 
large amplitude and very high in the potential hill. The total energy
{\em density} is non-perturbatively 
large. Classically the zero mode will undergo oscillatory behavior
between the two classical turning 
points, of very large amplitude and the dynamics will probe both
broken symmetry states. Even 
at the classical level the symmetry is respected by the dynamics in
the sense that the time evolution 
of the zero mode samples equally both vacua.

This situation is reminiscent of finite temperature in which case the
energy density is finite and above 
a critical temperature the ensemble averages sample both tree level
vacua with equal probability 
thus restoring the symmetry. In the  dynamical case, the ``symmetry
restoration'' is just a consequence 
of the fact that there is a very large energy density in the initial
state, much larger than the top of the 
tree level potential, thus under the dynamical evolution the system
samples both vacua equally.  

Thus the criterion for symmetry restoration  when the
tree level potential allows 
for broken symmetry states is that the energy density in the initial
state be larger than the top of the 
tree level potential. That is when the amplitude of the zero mode is
such that $ V(\eta_0) > V(0) $.  
In this case the dynamics will be very similar to the unbroken
symmetry case, the amplitude of the 
zero mode will damp out, transferring energy to the quantum
fluctuations via parametric amplification, 
but asymptotically oscillating around zero with a fairly large amplitude.

\section{Conclusions} 

We here answered the points raised by KLS in hep-ph/9608341. We refer
to our paper \cite{big} where in  
over 50 pages of text plus twenty figures, we  provide a
detailed and exhaustive analysis of the non-equilibrium issues in preheating
both in the unbroken and broken symmetry phases, in Minkowski space-time,
within a well defined non-perturbative, consistent, renormalizable and energy
conserving scheme.  We have given all the necessary technical details
including the error estimates in the numerical study and included two lengthy
appendices with the necessary details for the reader to follow all and every
step of our analysis.

\end{document}